\documentstyle{mn}

\newcommand{\ad}{$^o\,$}
\newcommand{\as}{$^{\prime\prime}~$}
\newcommand{\am}{$^{\prime}~$}

\newcommand{\etal}{ et al. }

\newcommand{\co}{\rm}
\def\ltsima{$\; \buildrel < \over \sim \;$}
\def\lsim{\lower.5ex\hbox{\ltsima}}
\input psfig.tex
\title[12\,min pulsations from 1WGA~J1958.2+3232]
{The discovery of  12\,min X--ray pulsations from 1WGA~J1958.2+3232}
\author[G. L. Israel et al.] 
{G. L. Israel,$^{1,}$\thanks{Affiliated to ICRA.} 
L. Angelini,$^{2,3}$ 
S. Campana,$^{4, }$\raisebox{2mm}{$\star$} 
P. Giommi,$^5$
L. Stella$^{1, }$\raisebox{2mm}{$\star$} and
\newauthor
N. E. White\,$^2$ \\
$^1$Osservatorio Astronomico di Roma, Vicolo dell'Osservatorio 2, 
I--00040 Monteporzio Catone (Roma), 
Italy\\ e--mail: gianluca and stella@coma.mporzio.astro.it\\
$^2$ Laboratory for High Energy Astrophysics, Code 662, 
NASA -- Goddard Space Flight Center, Greenbelt, MD 20771, USA\\
e--mail: angelini@lheavx.gsfc.nasa.gov / white@adhoc.gsfc.nasa.gov\\
$^3$Universities Space Research Association \\
$^4$Osservatorio Astronomico di Brera, Via E. Bianchi 46, 
I--23807  Merate (Lecco), Italy\\ 
e--mail: campana@merate.mi.astro.it\\
$^5$SAX Science Data Center, ASI, Viale Regina Margherita 202, 
I--00198 Roma, Italy \\
e--mail: giommi@napa.sdc.asi.it\\}

\date{Received 1997 Sep\,. Accepted 1998 Feb}

\begin{document}

\label{firstpage}

\maketitle

\begin{abstract}
During a systematic search for periodic signals in a sample of ROSAT 
PSPC (0.1--2.4 keV) light curves, we discovered  {\co $\sim$ 12~min} large 
amplitude X--ray
pulsations in 1WGA~J1958.2+3232, an X--ray source which lies close to the 
galactic plane. The energy spectrum is well fit by a power law 
with a photon index of 0.8, corresponding to an X--ray flux level of 
$\sim 10^{-12}$~erg~cm$^{-2}$~s$^{-1}$. The source is probably a long 
period, low luminosity X--ray pulsar, similar to X Per, or an intermediate 
polar . 
\end{abstract}


\section{Introduction}

Recent ROSAT observations have substantially increased the number of 
known X--ray pulsators (XRPs; Hughes 1994; Dennerl, Haberl \& Pietsch 1995; 
McGrath \etal 1994; Israel \etal 1997a, 1997b) and cataclysmic variables 
(CVs; Reinsch \etal 1994; Shafter \etal 1995; Buckley \etal 1995; Haberl 
\& Motch 1995; Burwitz \etal 1996; Friedrich \etal 1996; Singh \etal 1996).

In this paper we report the discovery of highly significant 
pulsations at a period of {\co 12\,min} in the X--ray flux of 
1WGA\,J1958.2+3232, 
a serendipitous ROSAT PSPC source which lies in the galactic plane. We also 
discuss the possible nature of the compact object responsible for these 
pulsations.

\section{X--Ray Observations and Data Analysis}

The field of 1WGA~J1958.2+3232 was observed on 1993 May 4 12:48--13:49 UT
($\sim 3700$~s exposure) and May 5 9:13--10:26 UT (exposure of $\sim 4300$~s) 
with the PSPC in the focal plane of the 
X--ray telescope on board ROSAT. About 10 point--like X--ray sources were 
detected within the field of view centered on the supernova remnant 
K4--41 (G 69.7+1.0). 
{\co 1WGA~J1958.2+3232 is located $\sim$23$\arcmin$ away from the image 
center. As a result of the periodic wobble in the ROSAT 
pointing direction (Briel et al. 1994) the source is at times partially 
obscured by the central ring and by one of the radial ribs of the PSPC 
window support structure. 
Therefore we determined the source position from an image accumulated by 
using only those time intervals in which the source was far from the 
window support structure.
The source is located at RA = 19$^h$ 58$^m$ 13$^s$.8, 
DEC. = +32\ad 32\am 58\as.7 (equinox 2000).
The uncertainty radius of $\sim$ 30\as is dominated by the size 
of the Point Spread Function (PSF) $\sim$ 23\am away from the 
image centre, with partial obscuration due to the wobble playing a minor 
role (in the image obtained from the entire PSPC pointing the source 
position differed only by $\sim$ 3\as).}
The ROSAT event list of 1WGA~J1958.2+3232 were
extracted from a circle of $\sim$ 1.2\am radius (corresponding to 
an encircled energy of about 90\%) around the X--ray position. 
Out of the 480 photons contained in the circle we estimated that $\sim 120$ 
photons derive from the local background.  

\begin{figure*}
\centerline{\psfig{figure=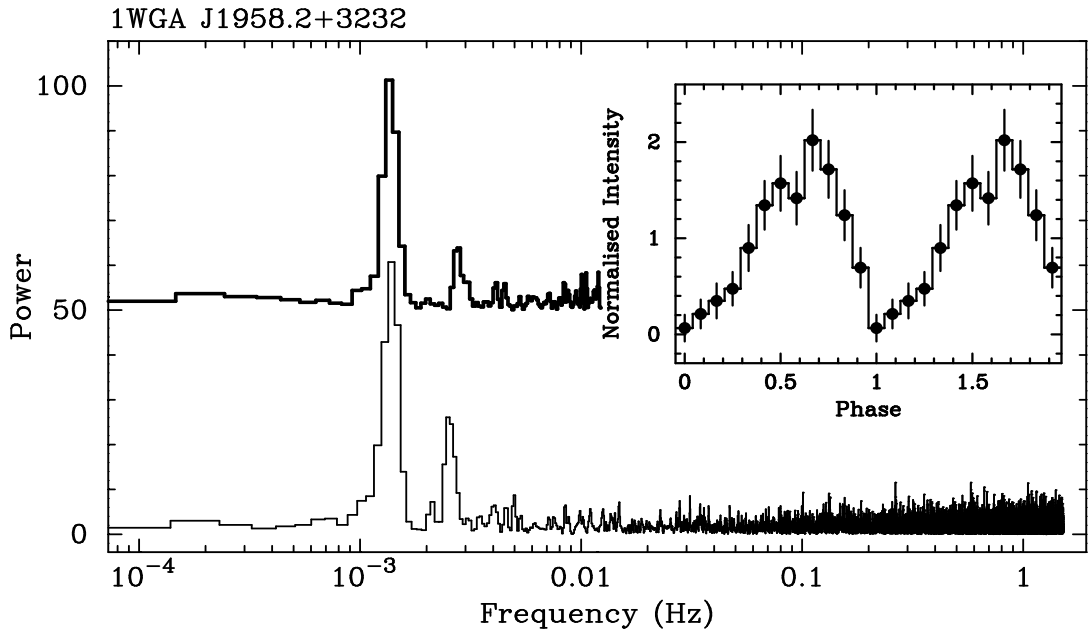,width=8cm,height=8.5cm} }
\caption{{\co Average power spectra of the original 0.1--2.4 keV ROSAT PSPC 
light curve of 1WGA~J1958.2+3232 obtained as described in the text (lower thin 
line). The counting statistics noise corresponds to a power of two. The thick 
upper curve gives the average power spectrum of the light curve corrected for 
the effects of the wobble (see text). The latter powers have been shifted 
by 50. The highest peak in both power spectra (centered around 0.00135 Hz), 
corresponds to the 12\,min modulation of 1WGA~J1958.2+3232. The corrected light 
curve folded at the 721\,s period is shown in the inner panel.}}
\end{figure*}

The May 1993 light curve of 1WGA~J1958.2+3232 was first analysed during 
a systematic study (Israel 1996) aimed at revealing periodicities 
in X--ray light curves of a sample of $\sim 23000$ X--ray sources selected 
from the White, Giommi \& Angelini catalog (1994; WGA catalog). The photon 
arrival times were corrected to the barycenter of the solar system.

In order to preserve 
the highest Fourier frequency resolution, a power spectrum was 
calculated over the entire observation time span ($\sim 0.9$~d). Four  
marginally significant peaks ($\geq 3.5\ \sigma$ over the entire spectrum)
were detected around a frequency of 1.4 $\times 10^{-3}$~ Hz. In order to 
minimise the windowing effects due to the 19.5 hr long gap between the 
May 4 and 5 observation intervals, we repeated 
the analysis by using two intervals ($\sim 5000$~s long), 
and calculated the average power spectrum from the power spectra 
obtained for each individual interval. {\co Several peaks 
were clearly seen around a frequency of 1.37 $\times 10^{-3}$ Hz 
(period of $\sim$ 740~s); the highest of these 
peaks has a significance of $\sim$ 8.5\ $\sigma$ over the entire sample of 
ROSAT light curves analysed (see Fig.~1 lower thin line).
The additional feature peaking at a frequency of 2.5 $\times 10^{-3}$~Hz
is consistent with the (partial) obscuration of the source caused by 
$402$~s wobble in the pointing direction. 
The 740~s modulation cannot represent the first sub--harmonics of the wobble 
period. Firstly, our systematic analysis of the power spectra of 
$\sim 23000$ light curves from  WGA sources (Israel 1996), shows 
that up to 10 higher harmonics of the wobble frequency are detected in 
a number of WGA sources. In no case, however, significant power at the 
first sub--harmonics of the wobble frequency is seen (neither it is clear 
how a modulation at 1/2 the wobble frequency could be introduced). 
Moreover the  1.37 $\times 10^{-3}$~Hz modulation of 1WGA~J1958.2+3232
is inconsistent with half the wobble frequency. The latter was confirmed 
to be at its usual value of $\nu_w$=2.49 $\times 10^{-3}$~Hz through the power 
spectrum analysis of the housekeeping data of the RA, DEC and roll angles of 
the pointing. The light curves of 1WGA~J1954.6+3222 and 1WGA~J1953.1+3251, two 
similarly bright X--ray sources in the field that were partially obscured 
by the central ring and the edge of the field itself, respectively, also showed 
significant peaks at 2.49 and 4.98 $\times 10^{-3}$~Hz. 

Based on the above results we conclude that the $\sim$ 12~min modulation 
($\nu_s$ = 1.37 $\times 10^{-3}$~Hz) is real and intrinsic to 1WGA~J1958.2+3232. 
Due to the presence of the wobble signal ($\nu_w$), sidelobes are expected  
in the power spectrum at frequencies 
$\nu_s\pm\nu_w$ (corresponding to $\sim$ 3.91 $\times 10^{-3}$~Hz 
and 1.15 $\times 10^{-3}$~Hz). 
A visual inspection of the power spectrum (lower thin line in Fig.\,1) shows 
that these are small if present at all. 
To approximately correct the source light curve for the effects caused by the 
wobble, we adopted the following procedure (see Fig.\,2): (a) we estimated the 
modulation introduced by the wobble by folding the light curve at the 402\,s 
period over 10 phase intervals (the mean intensity was normalised to one). (b) 
We accumulated a light curve with a binning time of 40.2\,s. (c) We devided 
each bin of the latter light curve by the corresponding bin of the light curve 
folded at the wobble period (see point a).  
The light curve obtained in this way is approximately corrected for the 
effects caused by the wobble (central panels of Fig.~2). 
By looking at the corrected light curve we note that the individual peaks of 
the source modulation at $\sim$ 12\,min are more clearly seen and the scatter 
of their amplitude is decreased compared to the original light curve.
The upper thick line in Fig.~1 shows the corresponding power spectrum of the 
corrected light curve. While the peak around a frequency of 1.37 
$\times 10^{-3}$~Hz is still present (a power of 53 for the average of two 
power spectra corresponding to a significance of 
$\sim$ 7.4 $\sigma$ over the entire sample of ROSAT light curves analysed), the 
peaks originating from the wobble disappear as expected. The 
second harmonics of the $\sim$ 12\,min signal is now clearly discernable around 
a frequency of $\sim$ 2.74 $\times 10^{-3}$~Hz (peak power of 14).}

Owing to poor statistics, 
a determination of the best period based on phase fitting proved unfeasible. 
Therefore we determined the best period in each of the two intervals by 
using a Rayleigh periodogram (cf. Leahy \etal 1983). The average of these 
periods gave {\co 721$\pm$14~s} (90\% uncertainties are used throughout this 
letter) {\co which is comparable to the Fourier resolution of the time span   
covered by the observation}. The modulation was 
nearly sinusoidal, with a pulse fraction (semiamplitude of modulation divided 
by the mean source count rate) of $\sim$ 80\% in the 0.1--2.4~keV band (see 
the insert in Fig.\,1). The arrival times of the pulse minima (that we 
adopted as phase 0) were determined to be JD 2449112.4838 $\pm$ 0.0004. 

\begin{figure}
\centerline{\psfig{figure=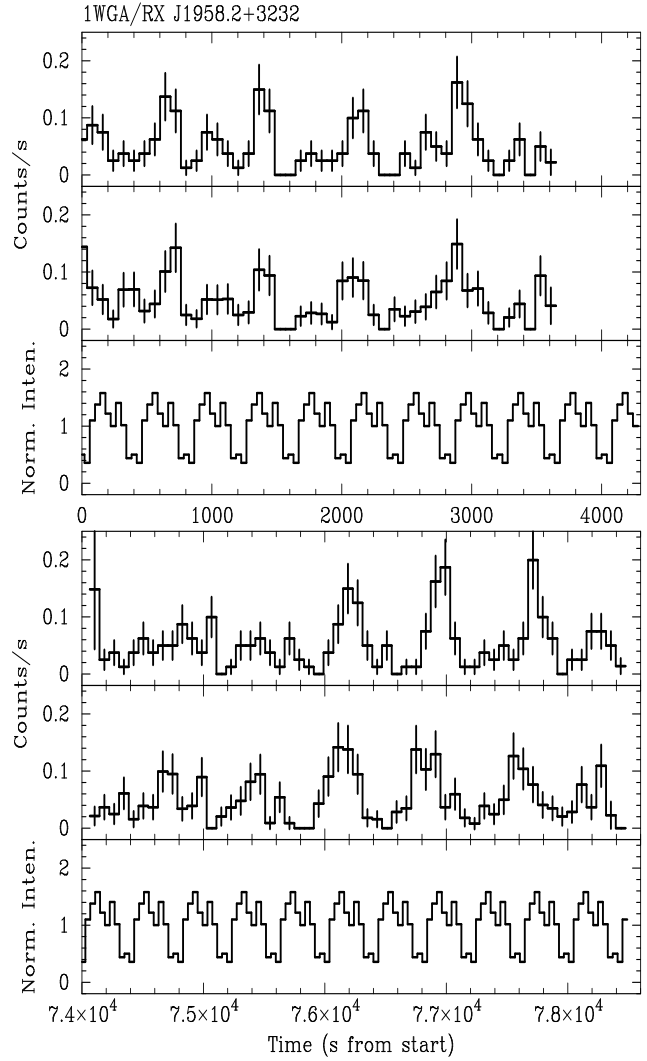,width=8cm,height=14cm} }
\caption{{\co ROSAT 0.1--2.4 keV PSPC X--ray light curve of 1WGA~J1958.2+3232 
obtained on 1993 May 4 (upper three panels) and May 5 (lower three panels). 
The first panel of each interval corresponds to the original light curve, 
while the central panel gives the corrected light curve (see text). In both 
cases the binning time is 80.4\,s. The third 
panel shows the light curve folded at the wobble period.}}
\end{figure}
 
The energy spectrum of 1WGA~J1958.2+3232 was well fit ($\chi^2$/d.o.f. 
= 13.7/16) by a simple power law (Fig.\,3a), a model often used  
to describe the soft X--ray spectrum of an X--ray pulsar. The best fit was 
obtained for a photon index of $\Gamma = 0.8\pm^{1.2}_{0.6}$ 
and a column density of $N_H = (6\pm^{24}_{5})\times 10^{20}~$~cm$^{-2}$. 
The corresponding 0.1--2.4~keV unabsorbed X--ray 
flux is $F_X \simeq 1.0 \times 10^{-12}$~erg~cm$^{-2}$~s$^{-1}$. 
The spectral parameters are poorly constrained as shown in Fig.\,3b 
where confidence contours are plotted in the $N_H$ -- $\Gamma$ plane. 
Note that the galactic hydrogen column in the direction of the source is  
$\sim 10^{22}$~cm$^{-2}$, while the  $3\ \sigma$ upper limit derived from the 
PSPC spectrum is  $4 \times 10^{21}$ cm$^{-2}$, {\it i.e.} a factor of 
$\sim 2.5$ lower. 
Among single component models a thermal bremsstrahlung gave 
also an acceptable fit. In particular, we obtained a $\chi^2$/d.o.f. = 
16.3/17 by fixing the plasma temperature $k\,T_{\rm br}$ at 
10 keV  
(an upper limit of 3 keV can be obtained from the data) in analogy with the 
spectrum inferred at higher energies for intermediate polars (IPs). 
In this case a column density of 
$N_H = (1.2\pm^{1.0}_{0.5})\times 10^{21}~$~cm$^{-2}$ was derived.

The position of 1WGA~J1958.2+3232 was included in two EXOSAT channel 
multiplier array (CMA; 0.05--2 kev) fields observed on 1983 October 6 
(25012 s effective exposure time) and 22 (10863 s). In both cases the source 
was not detected. The 0.1--2.4 keV flux upper limits (assuming a power law 
spectrum consistent with that derived from the ROSAT PSPC)  
are 1.2 $\times$ 10$^{-11}$ erg s$^{-1}$ cm$^{-2}$ and 1.3 $\times$ 10$^{-11}$ 
erg s$^{-1}$ cm$^{-2}$ for the October 6 and 22 observations, respectively.  

1WGA~J1958.2+3232 lies in a very crowded region close to the Galactic 
equator ({\it l}II=69$^o$.1, {\it b}II=1$^o$.7). About 15 stars with V $> 12$ 
mag are 
included in the current X--ray error circle (30\as radius; see Fig.\,4). 
The brightest star inside the error circle is 12.25 mag (RA = 19$^h$ 58$^m$ 
11$^s$.4, DEC = 32$^o$ 32$^{\prime}$ 58$^{\prime\prime}$.7, equinox 2000),  
while a 12.28 mag star is few arcsecs outside (see Fig.\,4) 
A visual inspection on the Palomar plates showed that the former star is 
bluer.  The other bright stars included in the error circle have  
V$\simeq$14.

\section{Discussion}

Based on available X--ray observations, it 
is not possible to unambiguously assess the nature of the compact object 
responsible for the X--ray pulsations observed in 1WGA~J1958.2+3232. 
The column density inferred from the ROSAT PSPC spectrum is substantially 
lower than the galactic value, suggesting a 
far shorter distance than the edge of the galaxy in the direction 
of the source ($\sim 15$~kpc).
Taking $d = 1$~kpc as an indicative value, 
the unabsorbed X--ray luminosity of the source is ${\rm L}_X = 1.2 \times 
10^{32}~(\frac{d}{1 {\rm kpc}})^2$ erg cm$^{-2}$ in the 0.1--2.4 keV band 
and ${\rm L}_X = 6 \times 10^{32}~(\frac{d}{1 {\rm kpc}})^2$ erg cm$^{-2}$ in the 
2--10 keV band (assuming a power law spectrum with $\Gamma = 0.8$). 

If 1WGA\,J1958.2+3232 contained 
an accreting magnetic neutron star, it would be 
one of a few XRPs known with rotation periods $> 500$ s. 
XRPs are mainly found in massive binary systems containing an O or B 
donor star. Their energy spectra are usually well modelled by relatively 
hard power law ($\Gamma \simeq 0-2$), with a cut--off at 
energies higher than 10 keV (White, Swank \& Holt 1983). 
Even though X--ray transient activity is common among systems with a Be--star 
primary, most long period ($\geq 100$~s) X--ray pulsars are persistent. 

\begin{figure}
\centerline{\psfig{figure=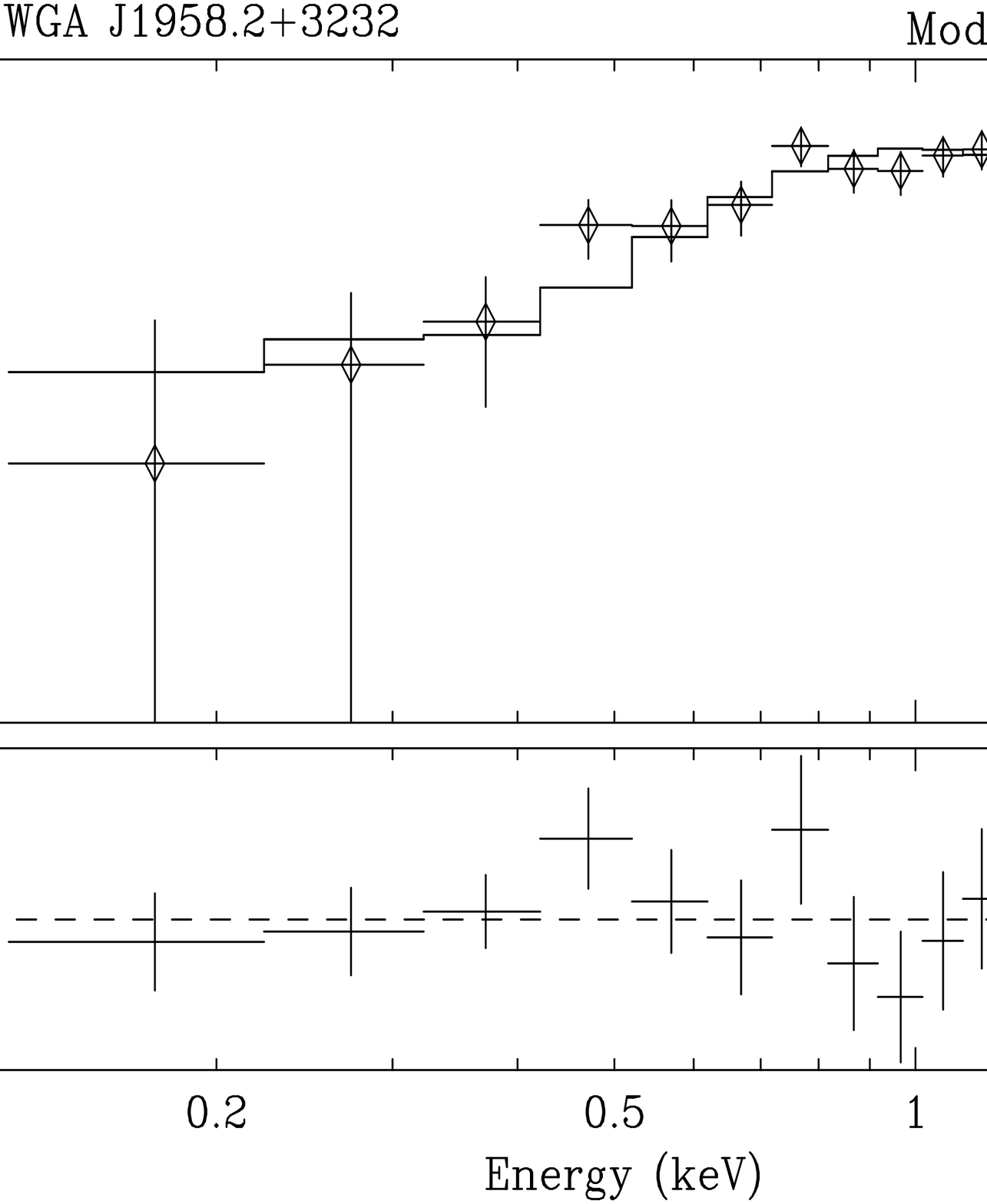,width=5cm,height=6cm} }
\centerline{\psfig{figure=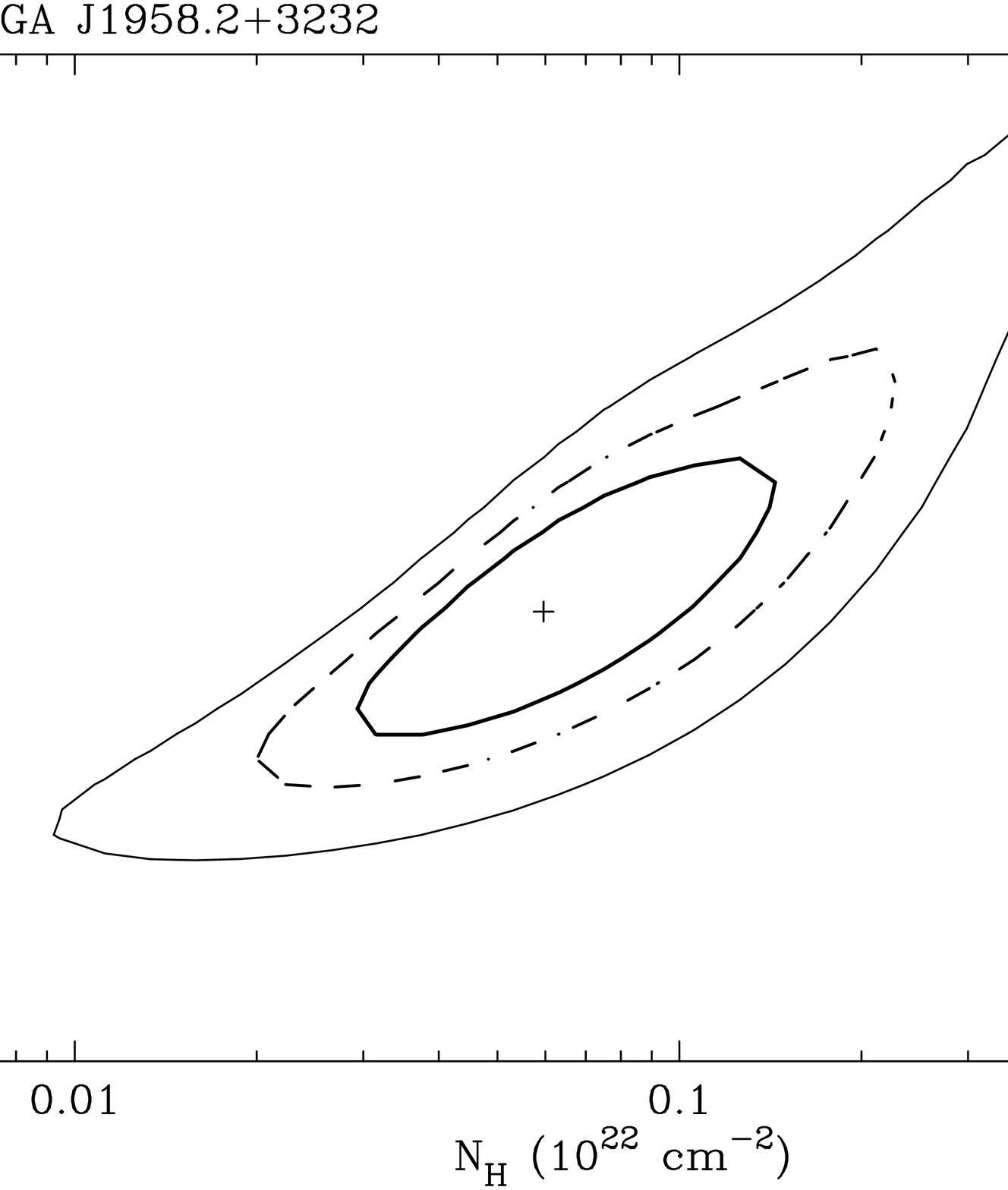,width=5cm,height=6cm}} 
\caption{ROSAT PSPC spectrum of 1WGA~J1958.2+3232. 
The best fit power law model is shown, together with the corresponding 
residuals. The lower panel gives the 1, 2 and $3\ \sigma$ confidence 
contours in the  $N_H$ -- $\Gamma$ plane. The cross indicates the best fit 
value.}
\end{figure}
    
Among these is X~Per, a low luminosity 
($L_X \sim 10^{33}-10^{34}$~erg~s$^{-1}$; 2--10 keV) 
835 s XRP with a peculiar 6.2 mag Be star companion
(Braes \& Miley 1972; White, Mason \& Sanford 1977; see Table 1). 
If 1WGA~J1958.2+3232 held a close resemblance to this system, then its 
distance would be in 3--5~kpc range, making one of the brighest stars 
(in particular the V=12.25 blue star) in the X--ray error circle the 
most likely candidate optical counterpart. 

Another system with similar properties is the 1455~s XRP RX~J0146+6121, 
a low galactic latitude X--ray binary with a 11.3 mag Be star primary
(White \etal 1987; Mereghetti, Stella \& De Nile, 1993; Israel, Mereghetti 
\& Stella 1994; {\co Hellier 1994; Haberl et al. 1997}). However, the low 
state X--ray luminosity of this system is likely much higher than that of 
1WGA~J1958.2+3232 ($\sim 2 \times 10^{35}$ erg s$^{-1}$ in the 2--20 keV 
energy band, for a distance of 2.5\,kpc; Hellier 1994; {\co $\sim$ (0.2--4.0) 
$\times 10^{35}$ erg s$^{-1}$ in the 0.5--10 keV energy band; Haberl \etal 
1997}).

\begin{figure}
\centerline{\psfig{figure=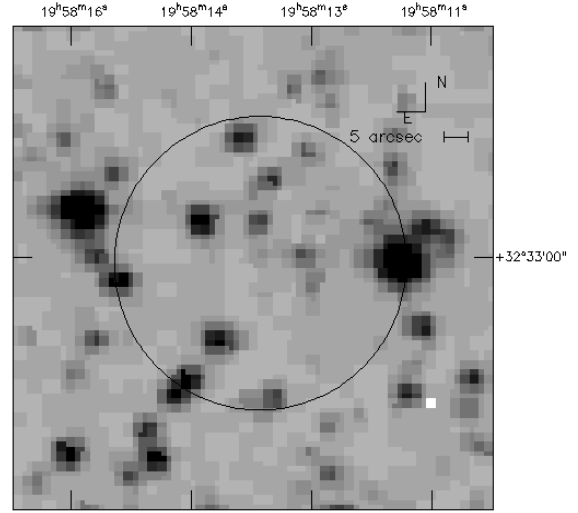,width=5cm,height=6.5cm} }
\vspace{5mm}
\caption{Digitalised Palomar chart of the region around 
1WGA~J1958.2+3232. The circle indicates the ROSAT PSPC error region.}
\end{figure}

Alternatively the {\co 12\,min} X--ray pulsations of 1WGA J1958.2+3232 might 
arise from polar cap accretion onto a magnetic rotating white 
dwarf. In this case 1WGA~J1958.2+3232 would 
belong to the intermediate polar class of cataclysmic variables. 
The energy spectra of IPs are, in most cases, well 
described by a thermal bremsstrahlung with a temperature 
$k\,T_{\rm br}$ of a few tens of keV. The 
white dwarf rotation periods are usually in the 5--30 min range, whereas the  
orbital periods have values of few hours.  

\begin{table*}
\begin{minipage}{130mm}
\begin{flushleft}
\begin{tabular}{c|cccccc}
\multicolumn{7}{c}{Table~1: Comparison of 1WGA~J1958.2+3232 with some XRPs 
and IPs observed by ROSAT}\\ 
\hline \hline 
Source & Period & Pulsed fr.  & $b$II        & PSPC Rate    & 
V--mag & 0.1--2.4 keV F$_X$ \\
       &s       &  \%&deg &ct\,s$^{-1}$&      & 
erg s$^{-1}$ cm$^{-2}$ \\ 
\hline 
1WGA~J1958.2+3232      & {\co 721}  & 80  & 1.7   & 0.045   & $> 12$ 
& 1.2 $\times 10^{-12}$ \\
\hline
X~Per            & 835  & 40  &--17   & 0.1--3.0 & 6.2, Be  
& 2--50 $\times 10^{-12}$\\
RX~J0146+6121          & 1412 & 80  & --0.8 & 0.2      & 11.3, Be  
& 5.6 $\times 10^{-12}$ \\    
\hline
RX~J1712.6--2414     & 1003   & 40  & 8.7   & 0.65     & 14.0     
& 9.5 $\times 10^{-12}$ \\   
RX~J0028.8+5917      & 313    & 40  & --3.5 & 0.6      & 14.4     
& 8.8 $\times 10^{-12}$ \\  
RX~J0153.3+7446      & 1414   & --  & 12.4  & 0.02     & ---      
& 3.4 $\times 10^{-13}$ \\ 
\hline
\end{tabular} 
\end{flushleft}
\end{minipage}
\end{table*}
Due to the relatively short distance ($<500$~pc) of most known IPs, 
their distribution covers a fairly wide range of galactic latitudes.  
However a sample of six intermediate polars 
located at low galactic latitude ($|$bII$|<$ 20$^{\rm o}$) 
has been recently identified in the 
ROSAT all--sky survey (Haberl \& Motch 1995). Among 
these, three (RX~J1712.6--2414, RX~J0153.3+7446 and 
RX~J0028.8+5917) are characterised by hard spectra 
consistent with those seen from classical IPs. 
Only the $V \simeq 14$ and $V \simeq 14.4$ optical counterparts 
of RX~J1712.6--2414 and RX~J0028.8+5917, respectively, have 
been studied in some detail (Motch \etal 1996; Buckley \etal 1995). 
The X--ray flux of these IPs is an order of magnitude higher than that of 
1WGA~J1958.2+3232, hinting to a factor of $\sim 3$ shorter distance. 
This analogy suggests that, if 1WGA~J1958.2+3232 is an IP, then its 
optical counterpart is likely in the 16--18~mag range.
Moreover note that the larger distance to 
1WGA~J1958.2+3232 would make its height above the galactic plane 
comparable to that of RX~J1712.6--2414 and RX~J0028.8+5917 (see Table 1).

Though unlikely, another possibility that cannot be rejected at present
is that the {\co 12~min} modulation of 1WGA~J1958.2+3232 corresponds to the 
orbital period of a low mass X--ray binary system. This would be only 
slightly longer that the shortest orbital period known, namely that of 
4U\,1820--303 (685~s;  Stella, Priedhorsky \& White 1987) 
However the pulsed fraction, in the case of 4U\,1820--303 
is only a few percents, much smaller than that of 1WGA~J1958.2+3232. 
A considerably higher X--ray orbital modulation might be expected if the 
accretion disk rim modulated the X--rays scattered in an extended accretion 
disk corona, while the central X--ray source is hidden from direct 
view due to a very high system inclination (see the case of 4U~1822--37; 
White \& Mason 1985). This might also account for the very low X--ray flux 
of  1WGA~J1958.2+3232 compared to that of (unobscured) low mass X--ray 
binaries.  

\section{Conclusions}
We discovered large amplitude {\co 12\,min} pulsations in the X--ray 
flux of 1WGA~J1958.2+3232. This periodicity likely originates from 
polar cap accretion onto a rotating magnetic compact star, either a 
white dwarf in an intermediate polar, or a neutron star in an X--ray 
binary system. In the latter case 1WGA~J1958.2+3232 
would be one of the very few XRPs known with a spin period of 
$> 500$ s and a low X--ray luminosity. 
If instead  1WGA~J1958.2+3232 hosts an accreting magnetic white 
dwarf rotation, it would likely represent a distant member of the 
class of low galactic latitude intermediate polars, recently discovered 
in the ROSAT all--sky survey. The additional possibility that the X--ray 
modulation arises from the orbital motion of a high inclination  
low mass X--ray binary with an accretion disk corona 
cannot be excluded at present. 

\section*{Acknowledgments}
This research has made use of data obtained through the High Energy
Astrophysics Science Archive Research Center (HEASARC), provided by 
NASA's Goddard Space Flight Center.
This work was partially supported through ASI grants. 
{\co The comments of an anonymous referee helped improving this letter.}

\vfill
\eject

\end{document}